\documentclass[%
 reprint,
 amsmath,amssymb,
 aps,
]{revtex4-1}

\usepackage{algpseudocode}
\usepackage{algorithm}
\usepackage{algorithmicx}
\usepackage{enumitem}
\usepackage{braket}
\usepackage{graphicx}
\usepackage{dcolumn}
\usepackage{bm}
\usepackage{animate}

\begin{document}


\title{\Large Quantum Algorithm to Solve a Maze :\\ \vspace{0.5mm} Converting the Maze Problem into a Search Problem}

\author{Debabrata Goswami}
\homepage{http://home.iitk.ac.in/$\sim$dgoswami/}
\affiliation{%
dgoswami@iitk.ac.in\\
 Indian Institute of Technology, Kanpur
}

\author{Niraj Kumar}
 \homepage{http://kumarniraj.weebly.com}
\affiliation{
kniraj@iitk.ac.in\\
Indian Institute of Technology, Kanpur
}


\begin{abstract}
We propose a different methodology towards approaching a Maze problem. We convert the problem into a Quantum Search Problem (QSP), and its solutions are sought for using the iterative Grover's Search Algorithm. Though the category of mazes we are looking at are of the NP complete class[1], we have redirected such a NP complete problem into a QSP. Our solution deals with two dimensional perfect mazes with no closed loops. We encode all possible individual paths from the starting point of the maze into a quantum register. A quantum fitness operator applied on the register encodes each individual with its fitness value. We propose an oracle design which marks all the individuals above a certain fitness value and use the Grover search algorithm to find one of the marked states. Iterating over this method, we approach towards the optimum solution. 
   
\end{abstract}

\maketitle
\section{Introduction}

In 1964, Gordon Moore, the co-founder of Intel, stated that the computing powers of computers would double about every 18 months and this is popularly known as the Moore's law [2]. If this were to continue to happen, then the number of transistors on the silicon chips would increase accordingly and the number of electrons in the transistor would drop as time goes on. In fact, Moore’s law has been more or less consistent, and presently the number of transistors in silicon chip for Core I7 Extreme Edition reaches 1.3 billion. This implies that only a handful of electrons make up a transistor, thus necessitating the need to work in Quantum regime. Such quantum computers have the potential to be the most powerful computational devices ever created, however, only a few practical applications of these have been devised effectively. One of the problems have been the design of quantum algorithm.\\

Quantum algorithms are the set of algorithms performed on the quantum bits employing the powerful concepts of quantum superposition and entanglement of the qubits. The interest in this domain was generated ever since physicists claimed that quantum computers use fewer resources (time and space) to perform certain computations as compared to classical computers (Deutsch $\&$ Jozsa algorithm) [3]. Shor's algorithm [4], for factoring numbers uses O($n^{2}log(n)log(log(n))$) operations as compared to classical factoring algorithm which uses $\Theta$($n^{1/3}log^{1/3}(n)$) operations and thus provides an exponential speed-up over the best known classical factoring algorithm. Also the Grover's search algorithm [5] of an unsorted data provides a quadratic speed-up, O($\sqrt{n}$) over the most efficient classical search algorithm O(n) . These have led to believe that QAs may have the potential of solving certain problems that seem intractable using Classical algorithms. \\

Solution to a general n $\times$ m Maze [1] is a NP complete problem and so it is an attractive problem to attack through quantum computing principles, which we present here. In context of the Maze problem, we define:\\
\begin{enumerate}[leftmargin=0.2in, topsep=0pt, partopsep=0pt]
\vspace{-1.5mm}
  \item {\it Individual}: One of the members of the population and a candidate for the solution to the Maze problem
  \item  {\it Individual Register}: The entire population containing a superposition of all individuals
  \item {\it Fitness}: Measure of how close a particular individual is to the optimal solution. Closer is the individual; higher is its fitness value.\\
\end{enumerate}

\vspace{-4mm}
\section{Overview of Strategy Used}

\begin{figure}[!ht]
\centering
\includegraphics[width=90mm]{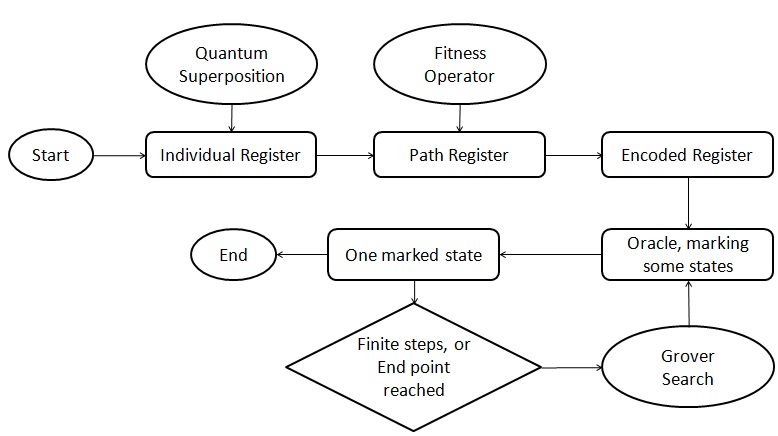}
\footnotesize
\caption{An outline of the method used in the paper to solve a Maze Problem.}
\label{overflow}
\end{figure}

\begin{enumerate}[leftmargin=0.2in, topsep=0pt, partopsep=0pt]

	\item First step involves the creation of a perfect square maze using the Recursive Backtracker Algorithm[6] \vspace{-1mm}
	\item Creation of individual register using quantum superposition storing the entire population. \vspace{-1mm}
	\item Classifying the individuals based upon their closeness to the optimum solution i.e. defining a fitness register storing the fitness values of individuals. \vspace{-1mm}
           \item Defining an Oracle which marks certain individuals having fitness values greater than a threshold fitness value.\vspace{-1mm}
           \item  Employing Grover's search algorithm in finding out one of the oracle's marked states.
           \item Iterating over the above two steps to find either the optimum individual or an individual nearing the optimum value.	\vspace{-1mm}

\end{enumerate} 
\vspace{2mm}
Now let us address the individual issues in the next few sections: 
\vspace{-2mm}
\section{Maze Creation}
The maze is a {\it m} $\times$ {\it m} two dimensional pathway where {\it m} is the number of rows and also the number of columns. The maze creation algorithm used in the paper is recursive backtracker algorithm [6]. \\
The concept of a pushdown stack is employed to implement this algorithm. Each grid in the maze is called a Room.
\\
\begin{enumerate}[leftmargin=0.2in, topsep=0pt, partopsep=0pt]

	\item An empty stack is defined initially. \vspace{-1mm}
	\item A random room is chosen and gets pushed onto the stack. This room becomes the starting room. \vspace{-1mm}
	\item Another room adjacent to the first room is chosen randomly, and the door is opened between the two rooms. This room is the current room. \vspace{-1.5mm}
           \item The starting room is pushed onto the stack.\vspace{-1.5mm}
           \item While there's at least one room left in the stack, the following steps are repeated: \\ 
                   \begin{enumerate}[leftmargin=0.2in, topsep=0pt, partopsep=0pt]
                   \vspace{-2.5mm}
                             \item If there are any unconnected rooms next to the current room: \\
                                      \begin{enumerate}[leftmargin=0.2in, topsep=0pt, partopsep=0pt]
                                      \vspace{-2.5mm}
                                      \item A random room is selected and the door between the two is open.
                                      \vspace{-1.5mm}
                                      \item The current room gets pushed onto the stack and the door between the two is opened.
                                      \vspace{-1.5mm}
                                      \item  Return to the top of loop. \\
                                      \end{enumerate}
                                      \vspace{-2.5mm} 
                             \item  If there are no unconnected rooms next to the current room: \\ 
                                       \begin{enumerate}[leftmargin=0.2in, topsep=0pt, partopsep=0pt]
                                      \vspace{-2.5mm}
                                      \item A room is popped off the stack and becomes the current room.
                                      \vspace{-1.5mm}
                                      \item We return to the top of the loop. \\
                                      \end{enumerate}
                                      \vspace{-2.5mm} 
                  \end{enumerate}
           \item When the stack is empty, the maze gets formed. \vspace{-1mm}
           \item Each room has one or more of the basis gates, North ({\it N}), East ({\it E}), South ({\it S}), West ({\it W}) gates. They are assigned as kets with values: $\Ket{{\it N}}$ = $\Ket{00}$, $\Ket{{\it E}}$ = $\Ket{01}$, $\Ket{{\it S}}$ = $\Ket{10}$, $\Ket{{\it W}}$ = $\Ket{11}$. \vspace{-1mm}
          \item For example, for a particular room $\Ket{grid[i][j]}$ (i = 0 to m-1, j = 0 to m-1), if $\Ket{{\it N}}$ and $\Ket{{\it E}}$ are open, then  $\Ket{grid[i][j]}$ = $\Ket{00}$ + $\Ket{01}$. \vspace{-1mm}
           \item Values for each grid is stored in the $\Ket{grid[i][j]}$ register.
	\vspace{-1mm}
\end{enumerate} 
\vspace{-2mm}
\section{Quantum Algorithm }
\subsection{Superposition}
Let us focus on the creation of Individual Register storing entire population [7]. \\
\begin{enumerate}[leftmargin=0.2in, topsep=0pt, partopsep=0pt]
\vspace{-2mm}
	\item For the basis states $\Ket{0}$ and  $\Ket{1}$ , the Hadamard gate on the basis states acts in the following manner: 
\begin{align}
 H\Ket{0} = \frac{1}{\sqrt{2}} ( \Ket{0} + \Ket{1} )
\end{align}
\vspace{-5mm}
\begin{align}
 H\Ket{1} = \frac{1}{\sqrt{2}} ( \Ket{0} - \Ket{1} )
\end{align} \vspace{-5mm}
	\item The four gates are represented as : \vspace{-1mm}
\begin{align}
 \Ket{{\it N}} = \Ket{00}, \Ket{{\it E}} = \Ket{01}, \Ket{{\it S}} = \Ket{10}, \Ket{{\it W}} = \Ket{11}
\end{align} \vspace{-7mm}

These kets are linearly independent  $\Braket{i|j}$ = 1 for i=j and 0 otherwise. i , j $\in$ $\{$ N,E,S,W $\}$ \vspace{-1.5mm}
	\item An operator U is defined as U = H $\otimes$ H \vspace{-1.5mm}
           \item Let the start and end point for the maze be ($i_{i}$ , $j_{i}$) and ($ i_{f}$ , $j_{f}$) respectively.\vspace{-1.5mm}
           \item  Each individual in the individual register has a length defined as n = 2 $\times$ ($i_{f}$ - $i_{i}$ + $j_{f}$ - $j_{i}$) .\\ The length is just taken at hand to ensure that a solution can be reached within that length of an individual. \vspace{-1.5mm}
           \item The input $\Ket{x_{0}}$ =  $\Ket{0}^{\otimes 2n}$ =  $\Ket{N}^{\otimes n}$ is passed to the operator $U^{\otimes n}$ to obtain the path register $\Ket{\Omega}$.  \vspace{-3mm}
           \begin{align}
\Ket{\Omega} = \frac{1}{2^{n}}\sum_{x \in \left \{ N,E,S,W \right \}^{n}}^{4^{n} -1}\Ket{x}
\end{align}
\vspace{-4mm}
\end{enumerate} 
\begin{figure}[!ht]
\centering
\includegraphics[width=85mm]{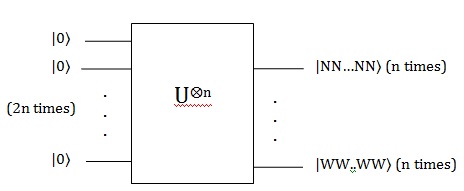}
\footnotesize
\label{overflow}
\end{figure}
\subsection{Fitness Evaluation of Individual Register}

\begin{enumerate}[leftmargin=0.2in, topsep=0pt, partopsep=0pt]
\vspace{-1mm}
  \item Under the fitness evaluation part, the {\it health} of a particular gene is calculated. It is a measure of the closeness of the individual to the optimum solution. Defining an operator for fitness evaluation of individual $\Ket{i}$ by operator {\it F} such that:
\begin{align}
 F\Ket{i} = \Ket{fit_{i}} 
\end{align}
\vspace{-7mm}
\begin{figure}[!ht]
\centering
\includegraphics[width=50mm]{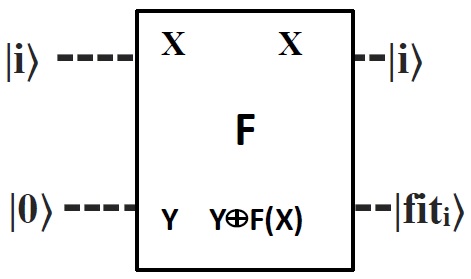}
\footnotesize
\caption{Figure depicting the Fitness operation on individual register.}
\label{overflow}
\end{figure}
\vspace{-2mm}
  \item  Let the starting point of the maze be ($i_{i}$,$j_{i}$) . Then for the individual $\Ket{i}$ = $\Ket{EWNNWE...S}$, the first step that’s possible is towards E(east) ($i_{i}$+1,$j_{i}$) . \vspace{-1mm}
  \item  Current Room: $i_{c}$ = $i_{i}$ , $j_{c}$ = $j_{i}$ and Next Room: $i_{n}$ = $i_{i}$+ 1 , $j_{n}$ = $j_{i}$.\vspace{-1mm}
  \item While the gate between Current Room and Next Room is open:
\begin{enumerate}[leftmargin=0.2in, topsep=0pt, partopsep=0pt]
\vspace{-1.5mm}
  \item Current Room = Next Room. 
  \item Next step is W. 
  \item If Current Room = End Room, then break out from the loop.
\item Return to top of the loop. 
\end{enumerate}
\item $\Ket{fit_{i}} = 2^{m} - dist(End Room - Current Room) = (i_{f} - i_{c})^{2} + (j_{f} - j_{c})^{2}$ \vspace{-1mm}
\item Fitness = $\Ket{fit_{i}}$ is an m-bit value. \\
\end{enumerate}
\begin{figure}[!ht]
\centering
\includegraphics[width=90mm]{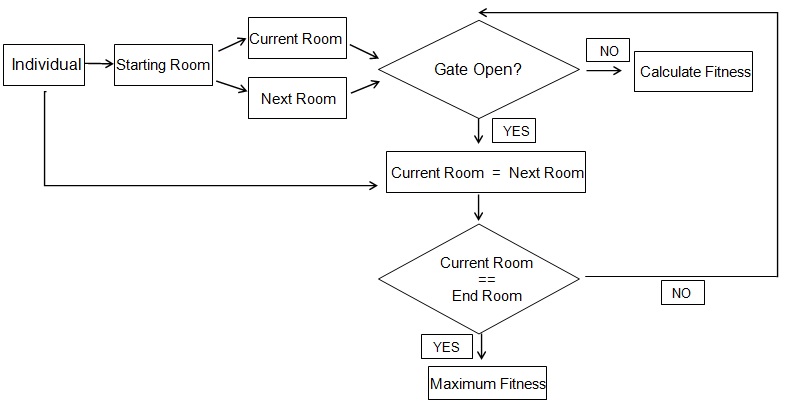}
\footnotesize
\caption{Flowchart depicting the functionality of  Fitness operator.}
\label{overflow}
\end{figure}

\subsection{Fitness Evaluation of Path Register}

The fitness evaluation discussed above was just for a single individual. But the fitness evaluation is needed for entire population. Quantum Parallelism is an important tool in quantum computation which enables to carry out fitness evaluation in just one step. The transformation is defined by the map: 
 \begin{align}
\Ket{\Psi} = F\Ket{\Omega} = \frac{1}{2^{n}}\sum_{x \in \left \{ N,E,S,W \right \}^{n}}^{4^{n} -1}\Ket{x} \otimes \Ket{fit_{x}}
\end{align}
\vspace{-4mm}
\begin{figure}[!ht]
\centering
\includegraphics[width=50mm]{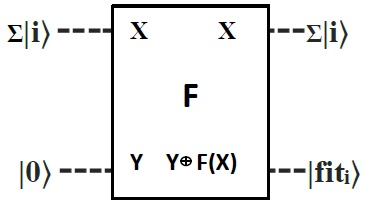}
\footnotesize
\caption{Figure depicting the Fitness operation on path register.}
\label{overflow}
\end{figure}
\vspace{-7mm}
\subsection{Finding the maximum}
Now that we have $\Ket{\Psi}$ containing the superposition of all individuals with their corresponding fitness values, the only task that remains is of finding the maximum fitness state. But, since the number of individual kets in the superposition ket $\Ket{\Psi}$ is O($4^{n}$), where {\it n} is the individual length, so finding the maximum fitness state directly via Grover’s algorithm is not feasible. \vspace{2mm} \\ 
Instead, a highly fit individual, one that is close to the optimum solution, is sought for after a finite number of iterations. An Oracle structure is defined which marks the states having fitness values greater than the cutoff fitness and then the Grover’s Algorithm is employed to find one of those marked states. This step is carried out for finite number of iterations and then the highly fit individual is looked up.

\subsection{Oracle Structure}
Quantum Search Algorithm employs the Grover operator whose first step is defining an Oracle to mark certain individuals based on the fitness criteria [5]. The steps employed in defining an Oracle are these: \\
\begin{enumerate}[leftmargin=0.2in, topsep=0pt, partopsep=0pt]

	\item A random fitness value {\it cutoff} = $\Ket{fit_{x}}$ is chosen from the set of fitness values. \vspace{-1mm}
	\item An oracle O is designed such that it marks all the kets in the superposition ket $\Ket{\Psi}$ that have fitness values greater than {\it cutoff}. \vspace{-1mm}
	\item The function F operates such that it flips those input states whose fitness is equal to or less than the {\it cutoff} fitness, and retains those states whose fitness is above the {\it cutoff} fitness. \vspace{-1mm}
\begin{align}
\Ket{x} \otimes \Ket{fit_{x}} \rightarrow -1^{f(fit_{x})}\Ket{x} \otimes \Ket{fit_{x}}
\end{align}
\vspace{-7mm}
\begin{align*}
f(fit_{x}) &= 1 , if fit_{x} > {\it cutoff} \\
                &= -1 , otherwise 
\end{align*}
 \vspace{-7mm}
           \item Defining an Oracle which marks certain individuals having fitness values greater than a threshold fitness value.\\
\end{enumerate}

\begin{figure}[!ht]
\centering
\includegraphics[width=90mm]{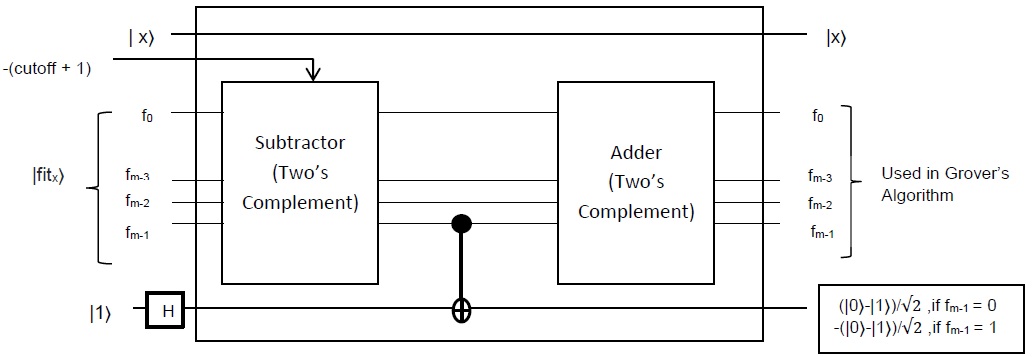}
\footnotesize
\caption{Figure depicting the Fitness operation on path register.}
\label{overflow}
\end{figure} 
\vspace{-5mm}
\subsection{Grover's Iteration to find the marked states}
After defining the Oracle structure, Grover Operator is used to find one of the marked states. Let {\it j} number of fitness states are marked by the oracle which satisfy f($fit_{x}$) = 1. 
 \begin{align}
\Ket{\Psi} &= F\Ket{\Omega} = \frac{1}{2^{n}}\sum_{x \in \left \{ N,E,S,W \right \}^{n}}^{4^{n} -1}\Ket{x} \otimes \Ket{fit_{x}} \\
                &=   \frac{l}{2^{n}}\sum_{O_{k}=1}\Ket{x} \otimes \Ket{fit_{x}} + \frac{\sqrt{4^{n}-l}}{2^{n}}\sum_{O_{k}=-1}\Ket{x} \otimes \Ket{fit_{x}} 
\end{align}
Thus the original register can be written into two different parts, ones for which $O_{k}$ = 1 (the marked ones) and ones for which $O_{k}$ = -1. Also, because multiple individuals may correspond to a single fitness values, hence the value {\it m} would in normal case differ from {\it l}. And in the above division, it is the {\it m} combined state which have $O_{k}$ = 1 and the rest $\sqrt{4^{n}-l}$ states have $O_{k}$ = -1. \\

Now in the next iteration, the {\it cutoff} value is updated by one of the marked state found by the Grover's Algorithm.In subsequent finite number of iterations, a very high fitness state can be found out i.e. approaching the optimum solution. \\

\subsection{Summary of the Algorithm}
\begin{enumerate}[leftmargin=0.2in, topsep=0pt, partopsep=0pt]
  \item A path register $\Ket{\Omega}$ storing all possible individual states is created. \vspace{1mm}
  \item Fitness operator F is applied and $\Ket{\Psi}$ state is created with $\Ket{\Psi}$ containing the superposition of all possible individuals with their fitness values. \vspace{-2mm}
\item (The solution is to be found in m steps) \vspace{1mm}
\begin{algorithmic}
   \State i = 0  $\#$ Start of Iteration
  \While{$i < m$}
    \State {\it cutoff} = $\Ket{fit_{x_{0}}}$.\vspace{0.5mm} 
    \State  Oracle $O_{i}$ marks all $\Ket{fit_{x}} > {\it cutoff}$. \vspace{0.5mm}  
    \State Grover search to find one marked state $\Ket{fit_{k}}$. \vspace{0.5mm} 
    \State {\it cutoff} = $\Ket{fit_{k}}$ . \vspace{0.5mm} 
    \State i++  \vspace{0.5mm} 
  \EndWhile 
\end{algorithmic}
\item At the end of m steps, either the maximum fit gene is obtained or we obtain a highly fit gene close to the optimum solution.
\item On measuring the fitness ket, the individual (or individuals) possessing the given fitness value is obtained because of the entanglement of fitness ket with the individual ket.
\end{enumerate}

\section{Complexity Assessment}
It is known that if {\it k} is the length of search space, the complexity of Grover’s algorithm is given by O($\sqrt{k}$). But here the population is exponential i.e. {\it k} = $4^{n}$. So the Grover’s algorithm cannot be applied directly to find the optimum solution in polynomial number of steps. Instead, we have defined the oracle and employed the iterative grover search algorithm to look for a high fitness solution in polynomial steps. The algorithm is designed to give a better(higher fitness) solution with each increasing iteration. Hence within {\it m} finite steps, it is not necessary that the best solution be reached, but a highly optimum solution can still be reached using this algorithm. 

\section{Conclusion}
Our main focus was on the conversion of a perfect maze into a quantum search problem and to approach towards the optimum solution in finite number of steps using iterative Grover search algorithm. The fitness gate is defined to calculate the ftness of each individuals, which is a criteria in selecting the optimum individuals. An Oracle structure has been proposed to mark certain individual states having fitness levels above the {\it cutoff} value. Grover's operator iterates subsequently, and with each iteration, the
closeness to the optimum solution increases. In the first iteration, if {\it m} individuals have fitness levels above the {\it cutoff}, an O($\sqrt{\frac{n}{m}}$) steps are used in the first iteration, with {\it n} being the total number of individuals. Next iteration is performed on the previous marked states, and so on, until the end of iteration is reached, or an optimum solution is found out. Grovers algorithm provides only the quadratic speedup, hence a solution is not guaranteed in the polynomial iteration steps. Subsequent search algorithms with more speedup would boost the approach towards the optimum solution.\\

The only limitation lies with the capability of the Grover's algorithm. Subsequent search algorithms, with power more than that of the Grover's algorithm, would boost the approach towards the optimum solution. A point to also note is that, if in future, the search algorithms are found with exponential speed up over the classical algorithms, then the NP-complete problem would have a solution in polynomial steps.

\end{document}